# Machine-learning potential for phonon transport in AlN with defects in multiple charge states

Ying Dou[1], Koji Shimizu[2,3], Jesús Carrete[4], Hiroshi Fujioka[1], and Satoshi Watanabe[2]

[1]Institute of Industrial Science, The University of Tokyo, 4-6-1 Komaba, Meguro, Tokyo 153-8505, Japan

[2]Department of Materials Engineering, The University of Tokyo, 7-3-1 Hongo, Bunkyo-ku, Tokyo 113-8656, Japan

[3]Research Center for Computational Design of Advanced Functional Materials (CD-FMat), National Institute of Advanced Industrial Science and Technology (AIST), Tsukuba Central 2, 1-1-1 Umezono, Tsukuba, Ibaraki 305-8568, Japan

[4]Instituto de Nanociencia y Materiales de Aragón (INMA), CSIC-Universidad de Zaragoza, Zaragoza 50009, Spain

## Abstract

Understanding phonon transport properties in defect-laden AlN is important for their device applications. Here, we construct a machine-learning potential to describe phonon transport with *ab initio* accuracy in pristine and defect-laden AlN, following the template of Behler-Parrinello-type neural network potentials (NNPs) but extending them to consider multiple charge states of defects. The high accuracy of our NNP in predicting second- and third-order interatomic force constants is demonstrated through calculations of phonon bands, three-phonon anharmonic, phonon-isotope and phonon-defect scattering rates, and thermal conductivity. In particular, our NNP accurately describes the difference in phonon-related properties among various native defects and among different charge states of the defects. They reveal that the phonon-defect scattering rates induced by $V_N^{3+}$ are the largest, followed by $V_{Al}^{3-}$, and that $V_N^{1+}$ is the least effective scatterer. This is further confirmed by the magnitude of the respective depressions of the thermal conductivity of AlN. Our findings reveal the significance of the contribution from structural distortions induced by defects to the elastic scattering rates. The present work shows the usefulness of our NNP scheme to cost-efficiently study phonon transport in partially disordered crystalline phases containing charged defects.

## I. INTRODUCTION

Aluminum nitride (AlN) belongs to a rare class of materials having both a large electronic bandgap (~6.1 eV) and a large thermal conductivity. Based on these properties, AlN plays a key role in solid-state lighting and modern power electronics. Owing to a small lattice mismatch, AlN is also widely used as a buffer during GaN growth in devices such as power high-electron mobility transistors (HEMTs). Semiconductor devices are sensitive to heat dissipation, and AlN-related semiconductors are no exception: especially under high-power and high-temperature operation conditions, heat dissipation is extremely important for AlN-related semiconductors to guarantee their performance and reliability.

Phonon-defect scattering is one of the main factors reducing the thermal conductivity in semiconductor materials, as defects with low formation energies are inevitably introduced during crystalline growth. Computationally, the strength of phonon scattering by defects is often underestimated by only taking into account small perturbations of the dynamical matrix coming from the defect mass difference [1, 2]. Specifically, in thermal conductivity calculations of AlN-related alloys and thin films, the behavior of acoustic-phonon scattering by impurities/defect has been mainly explained on the basis of the mass-difference effect [3, 4]. In studies using an analytical model based on a simplified Boltzmann transport equation (BTE) and the Debye approximation, Al vacancies were predicted as the strongest scatterers owing to their large mass difference with the host [5, 6]. Although vacancies, which can be modelled as created by disconnecting atoms from the lattice, do not introduce a mass perturbation per se, their structural distortions tend to be larger than those of substitutional defects, and this is sometimes modelled as an effective mass perturbation, such as six times the mass of the missing atom [7]. Those previous studies therefore relied on drastic simplifications for calculating the thermal conductivity of defect-laden crystalline systems by only considering a mass perturbation or replacing the effect of structural distortion with an effective mass perturbation.

In the recent decade, first-principles calculations of phonon transport properties have become possible thanks to the coupling of the BTE with density functional theory (DFT). In such calculations, both the effects of defect mass difference and structural distortion around the defect are included [7]. The effect of single substitutions, vacancies, coupled dopant-vacancy and dopant-dopant complexes, assumed to be in their most likely charged states, has been investigated in GaN [8]. The thermal conductivity of AlN laden with Al vacancies has also been studied [5]. Those calculation results agree well with the experimental measurements. However, the effects of defect charge states on phonon properties and thermal conductivities have not been extensively discussed yet. Very recently, some of us examined the phonon bands of AlN and GaN for pristine crystals and crystals with +1 or +3 nitrogen vacancies by DFT calculations, and revealed that the +3 charge state causes larger disturbances in phonon bands/density of states than the +1 one [9]. Based on this result, a difference in thermal conductivity must be seen between the +1 and +3 charge states due to the difference in interatomic force constants (IFCs), in spite of the fact that N vacancies in +1 and +3 charge states have the same mass-difference effect. Furthermore, consideration of charge state effects may overturn the previous speculations based on the defect mass difference. For instance, it would be worth re-examining whether the effect of the metal vacancy in III-nitrides is the greatest among native point defects.

In this study, $V_N^{1+}$, $V_N^{3+}$ and $V_{Al}^{3-}$, the most stable defects under Fermi level variance, are used to study the above issues. While *ab initio* calculations provide the highest accuracy, they usually incur high computational cost for large systems, particularly for phonon transport calculations of crystals with low symmetry. To overcome this problem, machine-learning potentials (MLPs) trained using *ab initio* calculation results are emerging as a promising alternative. Previous studies have verified that MLPs can describe phonon properties of defect-laden material systems with an accuracy comparable to *ab initio* calculations and much more reasonable cost [10-12]. However, as far as we know, MLPs have never been applied to examining thermal conductivities of defect-laden crystalline systems including multiple charge states. This challenge mainly stems from two difficulties:

1st, optimizing the fitting parameters of MLPs for multiple charged systems is difficult due to the nonunique output values for the same inputs on the structural information;

2nd, subtle phonon interactions in thermal conductivity calculations of charged-defect-laden crystalline systems require even higher prediction accuracy [11].

To overcome the first difficulty, some of us developed a modified high dimensional neural network potential (NNP), an extended Behler-Parrinello-type MLP [13], in a previous work [11]. In the present study, we demonstrate that this approach works well for predictions of phonon transport for multiple-charged defect-laden systems. Our approach achieves DFT-level accuracy in describing phonon transport for both pristine and defect-laden AlN including defects in multiple charge states. We find that $V_N^{3+}$ reduces the thermal conductivity more than $V_N^{1+}$. Moreover, our study overturns previous speculations based only on mass differences.

## II. METHODOLOGY

### A. Neural network potential architecture

The construction of the MLP in this study was based on the Behler-Parrinello-type NNP [13], but modified to add a system charge node ($Q_{sys}$) to the input layer [11]. It comprises a feed-forward neural network with two hidden layers. The local atomic features are described in the input nodes by means of radial and angular symmetry functions (SFs). $Q_{sys}$ is defined as constant value $c$ times a charge state of the target supercell $q_{sys}$ divided by its volume $V$: $Q_{sys} = c \frac{q_{sys}}{V}$. Here, $c = 100$ was used based on our previous experience [11]. The output node gives the atomic energy ($E_i$), which is expressed as

$$E_i = f^{out}\left[a_{0,1}^{out} + \sum_{k=1}^{k_0} a_{k,1}^{out} f^2 \left\{a_{0,k}^2 + \sum_{j=1}^{j_0} a_{j,k}^2 f^1 \left(a_{0,j}^1 + \sum_{\mu=1}^{\mu_0} a_{\mu,j}^1 G_i^\mu + a_{\mu_0+1,j}^1 Q_{sys}\right)\right\}\right], \quad (1)$$

where the $G_i^\mu$ are the SFs of atom $i$ (with an input node for each $\mu$) and $a_{l,m}^n$ denotes a weight parameter connecting the $n$th and precedent layers of the $l$th and $m$th nodes, respectively. $f^n$ corresponds to the activation function of the $n$th layer. The hyperbolic tangent was used for the first and second hidden layers and a linear function for the output layer. The total energy $E$ is the sum of $E_i$ over all atoms, $E = \sum_i E_i$. The forces are obtained as $F_{\alpha_i} = -\frac{\partial E}{\partial \alpha_i} = -\sum_\mu \frac{\partial E}{\partial G^\mu} \frac{\partial G^\mu}{\partial \alpha_i}$, where $\alpha = x, y, z$ atomic coordinates. Note that $Q_{sys}$ has no explicit contribution to the atomic forces.

**B. Training dataset generation**

To prepare training data for NNP, we first conducted first-principles calculations based on DFT using the Quantum Espresso software [14]. Calculations were performed using the plane-wave (PW) pseudopotential (PP) method and the generalized-gradient approximation (GGA) with the Perdew-Burke-Ernzerhof functional [15]. Ultrasoft PPs were taken from the GBRV library [16]. The Kohn-Sham wave functions and the charge density were expanded in PWs up to a kinetic-energy cutoff of 85 Ry and 850 Ry, respectively. We considered pristine AlN and AlN supercells containing one single point defect, which was $V_N^{1+}$, $V_N^{3+}$, $V_{Al}^{3-}$, $Al_i^{3-}$ and $Al_N^{3+}$, with several sizes consisting of 32 atoms to 128 atoms. For the purposes of this study, we concentrate on $V_N^{1+}$, $V_N^{3+}$, $V_{Al}^{3-}$ and do not discuss $Al_i^{3-}$ and $Al_N^{3+}$. The number of electrons was varied by hand for defects in $q$ charge states. In doing so, a jellium background charge of value $-q$ was also added to neutralize the systems. We used the lattice constants optimized for the pristine system, as well as those varied by ±1% to ±2% from the optimized ones for both pristine and defect-laden systems.

Besides the data directly obtained from first-principles calculations, data obtained by the following procedures were included: NNPs were roughly trained with the DFT outputs; then molecular dynamics (MD) were performed using Large-scale Atomic/Molecular Massively Parallel Simulator (LAMMPS) software [17, 18] with the roughly trained NNPs; Snapshot structures were extracted at certain intervals (5 to 30 steps) from the MD trajectory; these MD snapshot structures were used as inputs for DFT calculations and their outputs were added into training dataset.

To include structures having various structural features while restricting the total numbers of structures in the datasets used for training and validation, we performed principal component analysis (PCA) based on the SFs of each structure as performed in Ref. [19-21]. Local atomic features with dimension = number of SFs × N (number of atoms) were reduced to two-dimensional values, i.e., first and second principal components (PC1 and PC2). Structures having similar PC values are considered to possess similar atomic features. Conversely, we can expect to include structures with wide variety of local atomic environments by selecting structures so that their PC values are well scattered. We extracted structures according to this strategy (in practice, by using the procedure adopted in Ref. [19]), and in total, 18,558 structures consisting of 1,463,447 atoms were collected to train our final NNP. Among them, a randomly selected 10% were kept as the test dataset to monitor overfitting symptom and evaluate the prediction performance.

### C. NNP training

To optimize the weight parameters $\{a_{l,m}^n\}$, the following loss function was adopted:

$$\Gamma(\{a_{l,m}^n\}) = (1-\beta) \sum_{i=1}^{N_{\text{train}}} (E_i^{\text{NNP}} - E_i^{\text{DFT}})^2 + \beta \sum_{i=1}^{N_{\text{train}}} \left\{ \sum_{j=1}^{3n_i} (F_j^{i,\text{NNP}} - F_j^{i,\text{DFT}})^2 \right\},$$
(2)

where β controls the contribution ratio of energies and atomic forces. In the present study, β was set to 0.9. During the training process, the weight parameters were optimized to minimize the loss function using the limited-memory Broyden-Fletcher-Goldfarb-Shanno (l-BFGS) algorithm [22]. We generated 50 sets of initial weights, in which each weight parameter was set randomly within the range of [−1:1]. Then the set that provided the smallest root-mean-square error (RMSE) was adopted for the NNP training. The training was stopped when the RMSE of the training dataset became almost stationary while no symptom of overfitting was seen in the RMSE of the test dataset.

### D. Thermal conductivity calculations

The 2$^{\text{nd}}$ and 3$^{\text{rd}}$-order IFCs were calculated using the finite displacement method via the PHONOPY [23] and thirdorder.py [24] codes, respectively, with the default atomic displacement magnitudes. The thermal conductivities were calculated using the AlmaBTE package [25] with the following formula [24, 25]:

$$\kappa_{\mu\nu} = \frac{1}{k_B T^2 \Omega} \sum_j \int d\mathbf{q} \, n_{0_{j\mathbf{q}}}(n_{0_{j\mathbf{q}}} + 1)(\hbar\omega_{j\mathbf{q}})^2 v_{j\mathbf{q}}^\mu v_{j\mathbf{q}}^\nu \tau_{j\mathbf{q}}, \quad (3)$$

where $j$ denotes the phonon branch and q represents the wave vector, and $\mu$ and $\nu$ specify the Cartesian components of the $\kappa$ tensor. $T$ is the temperature and $\Omega$ is the unit cell volume. $\omega$, $v$, $n_0$, and $\tau$ are phonon-mode dependent quantities, corresponding to the phonon frequencies, group velocities, equilibrium occupancy, and relaxation time, respectively. Here, we considered the phonon-phonon anharmonic scattering ($\tau_{anh}^{-1}$), phonon-isotope scattering ($\tau_{iso}^{-1}$) and phonon-defect scattering ($\tau_{def}^{-1}$) as the contributions to the total mode-specific scattering rate. The $\tau_{def}^{-1}$ is calculated as [8, 26, 27]:

$$\tau_{j\mathbf{q},def}^{-1} = -\mathcal{X}_{def} \frac{A}{V_{def}} \frac{1}{\omega_{j\mathbf{q}}} \text{Im}\{\langle j\mathbf{q}|t|j\mathbf{q}\rangle\}, \quad (4)$$

where $\mathcal{X}_{def}$ denotes the number fraction of defects, $V_{def}$ represents the volume of a defect, and A is the volume used for normalizing $|j\mathbf{q}\rangle$. The **t** matrix in Eq. (4) is defined by $\mathbf{t} = (\mathbf{I} - \mathbf{V}\mathbf{g}^+)^{-1}\mathbf{V}$, where **I** is the identity matrix, $\mathbf{g}^+$ is the Green's function for the unperturbed crystal. **V** is the *ab initio* calculated perturbation matrix, which consists of both the mass variation contributions ($\mathbf{V}_M$) and the contribution of changes in force constants ($\mathbf{V}_K$) between the pristine and defect-laden structures [7, 8, 26, 27].

In this theory, $\tau_{anh}^{-1}$ and $\tau_{iso}^{-1}$ are calculated using 3$^{rd}$ and 2$^{nd}$-order IFCs of the pristine host, respectively, while $\tau_{def}^{-1}$ requires 2$^{nd}$-order IFCs of both the pristine host and the corresponding defect-laden structure. A non-analytical term correction (NAC) [28] was applied to the dynamical matrix of the pristine host, and further included into thermal conductivities of both pristine and defect-laden AlN. In both DFT and NNP predictions, NAC ingredients calculated using Quantum Espresso were used. Unless otherwise specified, a q-point mesh of 26×26×14 for all calculations of phonon mode scattering rates, and a grid of 18×18×10 q points for the Green's functions used for calculating $\tau_{def}^{-1}$ [8].

### III. RESULTS AND DISCUSSION

**A. NNP construction**

In our NNP construction, we used 8 and 24 types of radial and angular SFs, respectively, for each elemental combination (8×2 + 24×3 = 88) within a 7-Å cutoff distance (See Supplemental Material S1 for details of SFs and the cutoff function). We used one system charge node in the input layer, and 30 and 20 nodes in the first and second hidden layers, respectively. Thus, the NN architecture was [89-30-20-1]. Comparison between DFT calculations and the predictions using the constructed NNP regarding total energies and atomic forces is depicted in Figure 1. Both the total energies and atomic forces closely aligned along the diagonal lines, indicating that the constructed NNP accurately predicted all the structures and charge states considered. The RMSEs of the total energies and atomic forces of the proposed model were 3.55 meV/atom and 69.16 meV/Å for the training dataset, and 3.48 meV/atom and 68.99 meV/Å for the test dataset, respectively. The RMSEs for each type of systems also showed high prediction accuracy (See Supplemental Material S2). The PCA on the SFs revealed that the structural features of the relaxed defect-laden structures were in the range spanned by those of the training dataset, although those relaxed defect-laden structures were not used for training (See Supplemental Material S3). This confirmed that our training dataset has sufficient variety, thereby avoiding the lower prediction accuracy due to the lack of relaxed defect structural features seen in Ref. [11], where structures of the training dataset was created by molecular dynamics using the empirical potential.

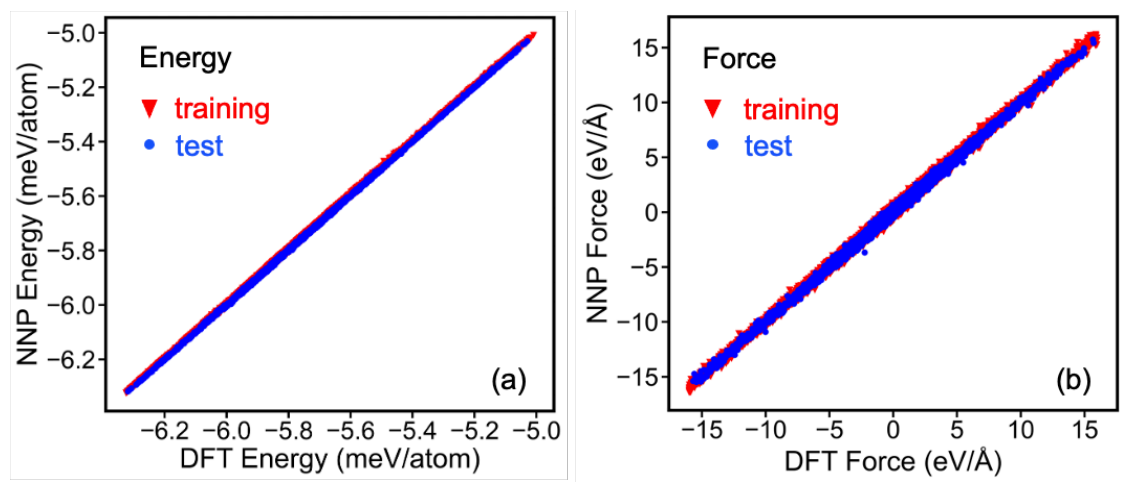

FIG.1. Comparison between DFT and NNP in respect of (a) total energies and (b) atomic forces. The red triangles and blue circles represent the training and test data, respectively.

**B. Relaxation of defect-laden structures**

To get a converged thermal conductivity, a large enough supercell size is necessary. The lighter computational cost of the NNP compared with DFT allowed us to run a stringent convergence test of $\kappa$ (See Supplemental Material S4) versus the supercell size. Based on this test, we decided to use a 4×4×3 supercell size consisting of 192 atoms for pristine AlN, or 191 for AlN including one vacancy, in the following study. We chose this supercell size considering the convergence on the cell size as well as the heavy computational cost of DFT calculations used for verification. Phonon transport properties were also confirmed to be converged when adopting this size (S4-1 in the Supplemental Material and Ref. [8]). It is noteworthy that this supercell size is larger than those involved in the NNP training dataset, and thus provides a particularly severe test for the predictive power of the machine-learning potential.

In Figure 2, we show the relaxed atomic configurations of the defect-laden AlN supercells obtained by DFT and NNP. The interatomic distances after relaxation, and their changes compared with the corresponding pristine values are listed in Table 1. Our NNP prediction of structural relaxations around defects agreed with DFT well. The structural distortions around $V_N^{1+}$ were the smallest. As expected, the structural distortions of atoms surrounding the defect were larger than those of the other atoms. Although Al atoms are heavier than N atoms, the structural distortions around $V_N^{3+}$ were larger than $V_{Al}^{3-}$.

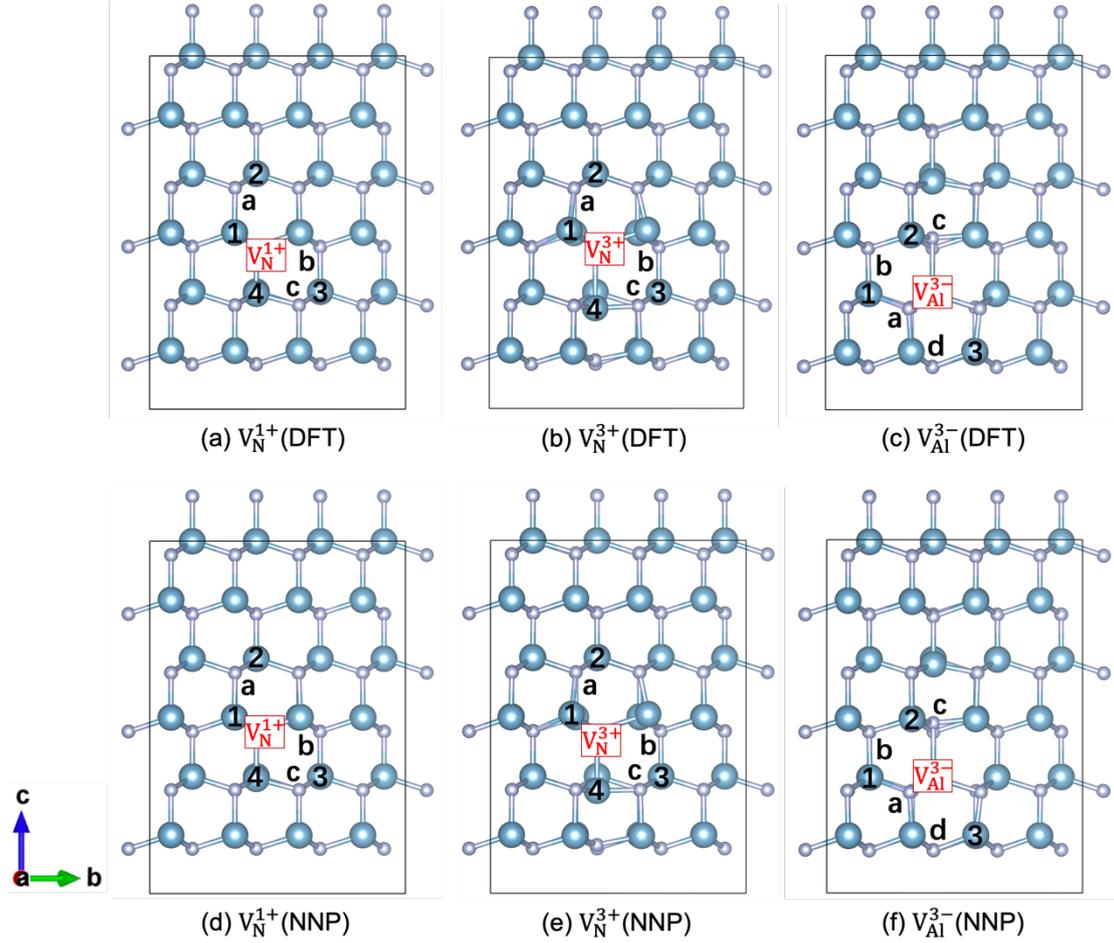

FIG. 2. Atomic configurations of optimized defect-laden structures obtained by (a-c) DFT and (d-f) NNP: (a, d) $V_N^{1+}$, (b, e) $V_N^{3+}$ and (c, f) $V_{Al}^{3-}$.

TABLE I. Distances between defect and surrounding Al and N, V-Al and V-N. The atom IDs are denoted in Fig. 2. The atomic distances in pristine AlN are the same in the predictions of NNP and DFT, which is denoted as $L_p$. The atomic distances predicted by DFT and NNP are denoted as $L_D$ and $L_N$, respectively.

| Defect | Body type | Distance (Å) | DFT $L_D$ | DFT $(L_D-L_p)/L_p$ | NNP $L_N$ | NNP $(L_N-L_p)/L_p$ | Pristine $L_p$ |
|---|---|---|---|---|---|---|---|
| $V_N^{1+}$ | V-Al | V-1 | 1.951 | 2.45% | 1.970 | 3.45% | 1.904 |
| | | V-2 | 3.119 | 0.35% | 3.112 | 0.11% | 3.108 |
| | | V-3 | 3.683 | 0.31% | 3.682 | 0.28% | 3.672 |
| | | V-4 | 2.009 | 4.84% | 2.027 | 5.82% | 1.916 |
| | V-N | V-a | 3.079 | -0.52% | 3.079 | -0.51% | 3.095 |
| | | V-b | 3.122 | -0.31% | 3.126 | -0.19% | 3.132 |
| | | V-c | 3.084 | -0.38% | 3.086 | -0.30% | 3.095 |
| $V_N^{3+}$ | V-Al | V-1 | 2.328 | 22.29% | 2.281 | 19.79% | |
| | | V-2 | 3.138 | 0.95% | 3.140 | 1.03% | |
| | | V-3 | 3.695 | 0.64% | 3.699 | 0.75% | |
| | | V-4 | 2.626 | 37.06% | 2.554 | 33.31% | |
| | V-N | V-a | 3.098 | 0.11% | 3.088 | -0.23% | |
| | | V-b | 3.142 | 0.31% | 3.132 | -0.01% | |
| | | V-c | 3.131 | 1.16% | 3.112 | 0.53% | |
| $V_{Al}^{3-}$ | V-Al | V-1 | 3.097 | -1.11% | 3.102 | -0.95% | 3.132 |
| | | V-2 | 3.075 | -0.64% | 3.063 | -1.04% | 3.095 |
| | | V-3 | 3.032 | -2.05% | 3.037 | -1.88% | 3.095 |
| | V-N | V-a | 2.106 | 10.62% | 2.117 | 11.17% | 1.904 |
| | | V-b | 3.687 | 0.43% | 3.692 | 0.56% | 3.672 |
| | | V-c | 2.361 | 23.24% | 2.308 | 20.49% | 1.916 |
| | | V-d | 3.131 | 0.73% | 3.130 | 0.72% | 3.108 |

### C. Phonon properties

Figure 3 depicts the comparisons of phonon bands between DFT and NNP for pristine AlN and AlN supercells including a point defect: $V_N^{1+}$, $V_N^{3+}$ and $V_N^{3-}$. The phonon bands of pristine AlN predicted by DFT agree well with those predicted by the NNP and the LO-TO phonon splitting is correctly described, as shown in Fig. 3(a). The optimal lattice constants of our AlN unit cell obtained by DFT and NNP are the same, a = 3.132 Å and c = 5.024 Å. These values are different from the experimentally measured ones, a = 3.11 Å and c = 4.98 Å [29], leading to differences in phonon bands. We therefore also employed the NNP to predict phonon bands using the experimental lattice constants. Compared with the optimal lattice constants, the experimental values of the lattice constants cause an upward shift of the phonon bands, which results in good agreement with X-ray and Raman data. These results highlight the high accuracy of our NNP.

The phonon bands of defect-laden AlN (including $V_N^{1+}$, $V_N^{3+}$ and $V_{Al}^{3-}$) predicted by NNP show good agreement with those extracted from DFT. Specifically, the features of band discontinuity, band splitting and broadening of bands caused by defects reported in Ref. [9] are well predicted by the NNP. The highest-lying branch shows a slight deviation from DFT predictions. However, this discrepancy was deemed acceptable given its negligible impact on the vibrational density of states [11] and on phonon transport.

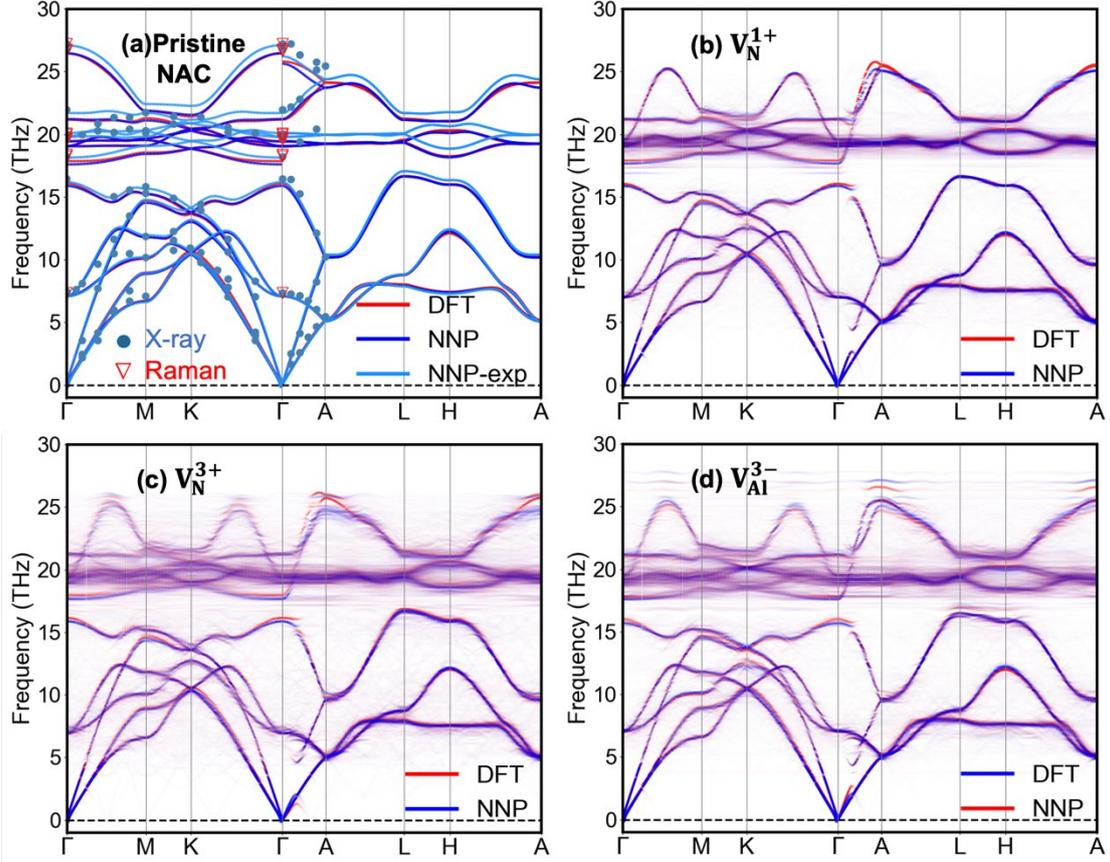

FIG. 3. Comparisons of phonon bands between NNP and DFT calculations for (a) pristine AlN, and AlN supercells with one point defect: (b) $V_N^{1+}$, (c) $V_N^{3+}$ and (d) $V_{Al}^{3-}$. For pristine AlN, the phonon bands predicted by NNP with experimental lattice constants are also shown. The X-ray [30] and Raman [31, 32] data were measured by experiments. The NAC is included for phonon bands of pristine AlN but not for those of defect-laden AlN. Note that the phonon bands in the Brillouin zone of the supercell are unfolded onto the primitive cell using a band unfolding package [33-36].

### D. Phonon scattering rate

As displayed in Figure 4, the three-phonon anharmonic, phonon-isotope and phonon-defect scattering rates predicted by NNP agree well with those predicted by DFT (See details of the parameters used in calculations of $V_K$ in the Supplemental Material - S5). Compared with $\tau_{def}^{-1}$ ($V_{Al}^{3-}$), the NNP prediction accuracies for $\tau_{def}^{-1}$ ($V_N^{1+}$) and $\tau_{def}^{-1}$ ($V_N^{3+}$) are a little worse but acceptable and reasonable, considering the difficulty of handling multiple charge states for the same defect species.

The overall importance of $\tau_{iso}^{-1}$ is minor. However, the isotope effect is more noticeable in the high-frequency region than in the low-frequency region. This is because the isotope effect of Al is smaller than that of N. Al contributes mostly to the low-frequency region, whereas N predominantly contributes to the high-frequency region.

The $\tau_{def}^{-1}$ is a function of defect concentration $D$, which in the present calculation is set to one defect per unit cell, just for convenience of comparisons. A phonon-defect scattering rate $\tau_{def}^{-1}(real)$ of a realistic defect concentration $D_{real}$ can be obtained using $\frac{D}{D_{real}} = \frac{\tau_{def}^{-1}}{\tau_{def}^{-1}(real)}$, where $D = \frac{1}{V_{unit}}$. The $D_{real}$ considered in this study covered a wide range of realistic defect concentrations from $10^{18}$ cm$^{-3}$ to $10^{20}$ cm$^{-3}$. In Figure 5, by employing NNP predictions, we compare $\tau_{anh}^{-1}$, $\tau_{iso}^{-1}$ and $\tau_{def}^{-1}$ with realistic defect concentrations at room temperature (RT, 293K) and 700K. When the defect concentration is low, the main contribution to the total mode-specific scattering rate comes from $\tau_{anh}^{-1}$. When the defect concentration is as high as $10^{20}$ cm$^{-3}$, the magnitude of $\tau_{def}^{-1}(real)$ is comparable to $\tau_{anh}^{-1}$. Among the three vacancy defects, the $\tau_{def}^{-1}$ was the smallest for $V_N^{1+}$, which can be easily understood from the smallest defect mass difference and negligible structural distortions around $V_N^{1+}$, as shown in Fig. 2(a, d). As for the comparisons of $\tau_{def}^{-1}$ between $V_N^{3+}$ and $V_{Al}^{3-}$, $\tau_{def}^{-1}(V_{Al}^{3-})$ was considered larger based on the previous theory using defect mass difference. However, our results contradict this speculation, as $\tau_{def}^{-1}(V_N^{3+})$ is slightly larger. This can be explained by the greater structural relaxation around $V_N^{3+}$ compared to $V_{Al}^{3-}$. This greater relaxation leads to larger changes in force constants, and thus creates a larger contribution to $\tau_{def}^{-1}$.

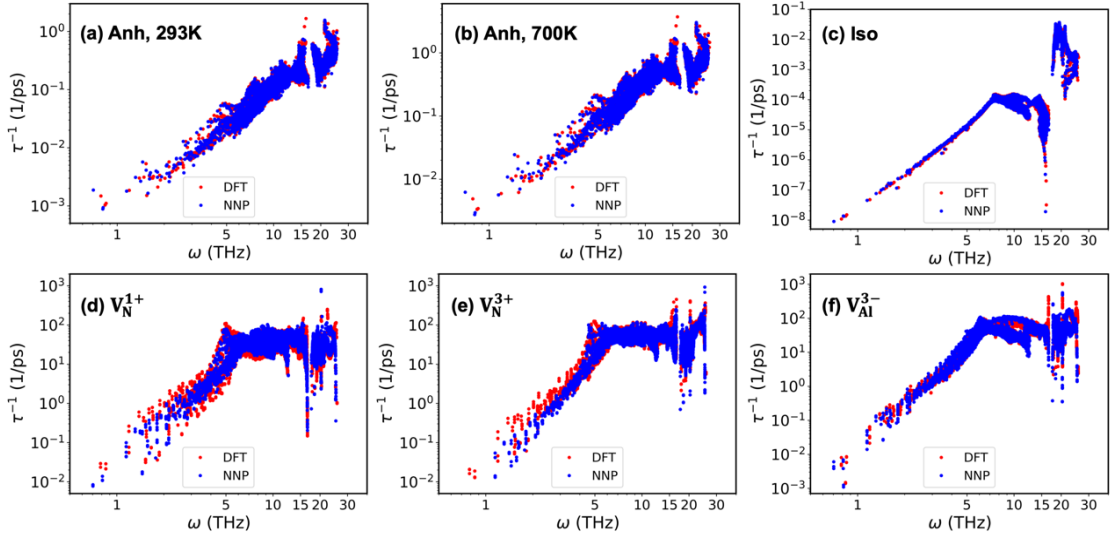

FIG. 4. Comparisons between NNP and DFT predictions for three-phonon anharmonic scattering rates at (a) 293K and (b) 700K, (c) phonon-isotope scattering rates, and phonon-defect scattering rates for (d) $V_N^{1+}$, (e) $V_N^{3+}$ and (f) $V_{Al}^{3-}$. Note that the phonon-defect scattering rate shown here were calculated using a defect concentration defined as one defect per unit cell.

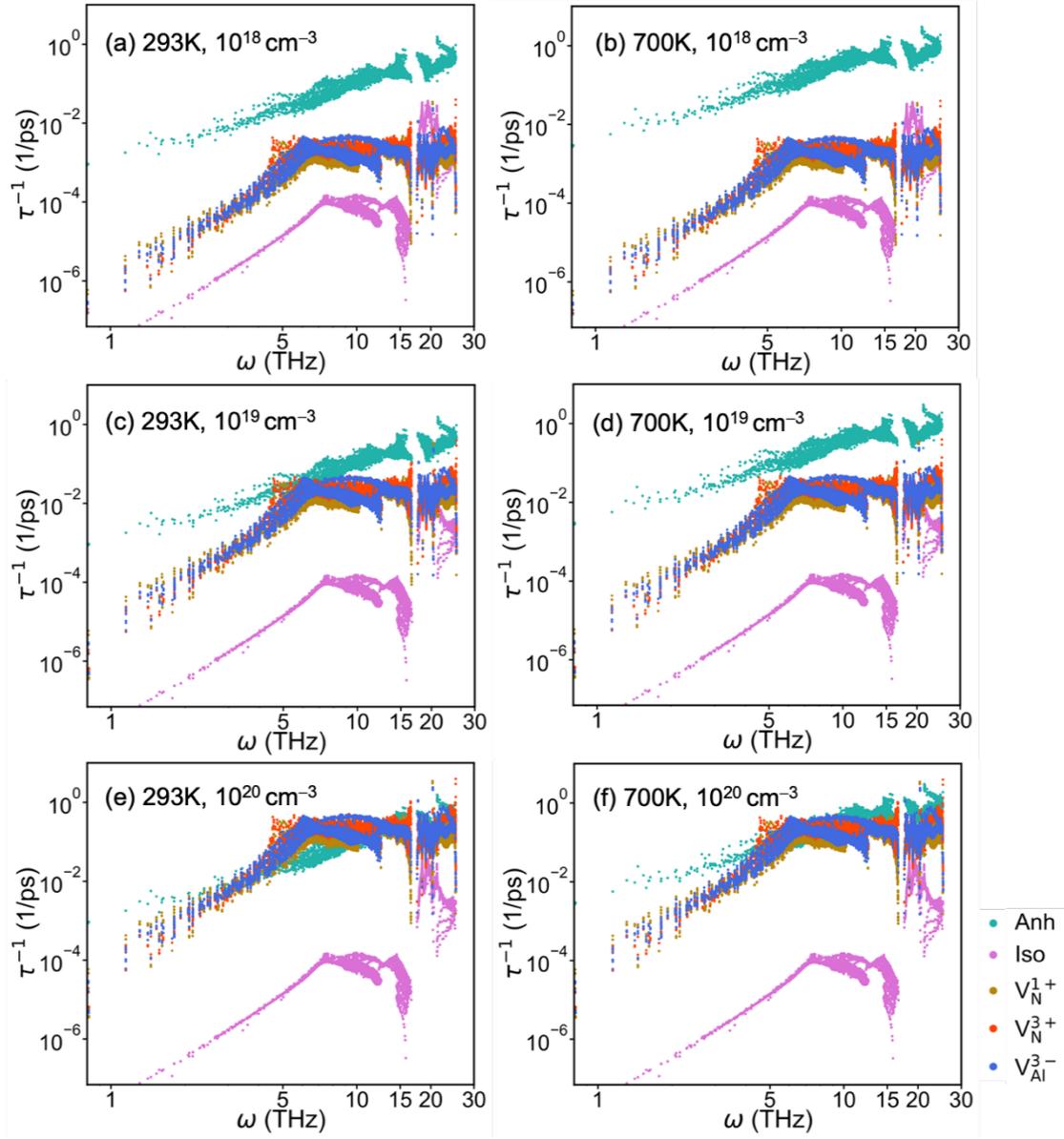

FIG. 5. Comparisons of phonon scattering rates among $\tau_{anh}^{-1}$, $\tau_{iso}^{-1}$ and $\tau_{def}^{-1}$ predicted by NNP at (a, c, e) 293K and (b, d, f) 700K. The defect concentrations were (a-b) $10^{18}$ cm$^{-3}$, (c-d) $10^{19}$ cm$^{-3}$ and (e-f) $10^{20}$ cm$^{-3}$.

### E. Thermal conductivity ($\kappa$)

In order to ensure the convergence of thermal conductivity, the 3$^{rd}$-order IFCs were considered up to the 10$^{th}$ nearest neighboring atoms with the help of predictions from the NNP and verification by DFT (See Supplemental Material S4-2). The average thermal conductivities of pristine and defect-laden AlN were predicted by NNP and DFT at temperatures from RT to 700K, as shown in Figure 6. Defect concentrations ranging from $10^{18}$ to $10^{20}$ cm$^{-3}$ were considered for each defect-laden AlN system. The in-plane and out-of-plane $\kappa$ for pristine and defect-laden AlN are shown in Figure S6 of the Supplemental Material. We found that the relative magnitude of in-plane and out-of-plane $\kappa$ depends on the q-point mesh. When the q-point mesh was sufficient to ensure a converged $\kappa$, the in-plane $\kappa$ is slightly larger than the out-of-plane $\kappa$ (Supplemental Material S4-3). Theoretical evaluation of $\kappa$ was found to be sensitive to the cutoff radius of the 3$^{rd}$-order IFC and the optimal lattice constants. On the other hand, the experimentally measured $\kappa$ depends on the growth techniques because of the difference in defect concentration and species in resultant samples, but precise assessment of defect concentration and species for each sample is usually difficult. Considering the above two points, instead of paying attention to the comparisons between our results and previous studies, we focused on the comparisons of phonon properties between DFT and NNP, and between pristine and defect-laden AlN using the same calculation settings in our study.

The good agreement with DFT calculations on isotropic thermal conductivities for pristine AlN and defect-laden AlN confirmed the accurate and promising prediction power of our NNP. The average $\kappa$ of pure AlN including three-phonon anharmonic scattering was 256.93 (253.72) Wm$^{-1}$K$^{-1}$ using NNP (DFT) at RT, while that of natural AlN additionally including isotopic effects was 256.56 (253.34) Wm$^{-1}$K$^{-1}$ using NNP (DFT). Thus, the isotope effect in AlN is negligible. Since the phonon bands of pristine AlN displayed clear differences when using the theoretically optimal and experimental lattice constants, $\kappa$ using the experimental lattice constants was also predicted by NNP, yielding values of 308.60 Wm$^{-1}$K$^{-1}$ and 276.83 Wm$^{-1}$K$^{-1}$ along the in-plane and out-of-plane directions at RT, respectively. The compression stress when using the experimental lattice constants thus leads to the enhancement of $\kappa$.

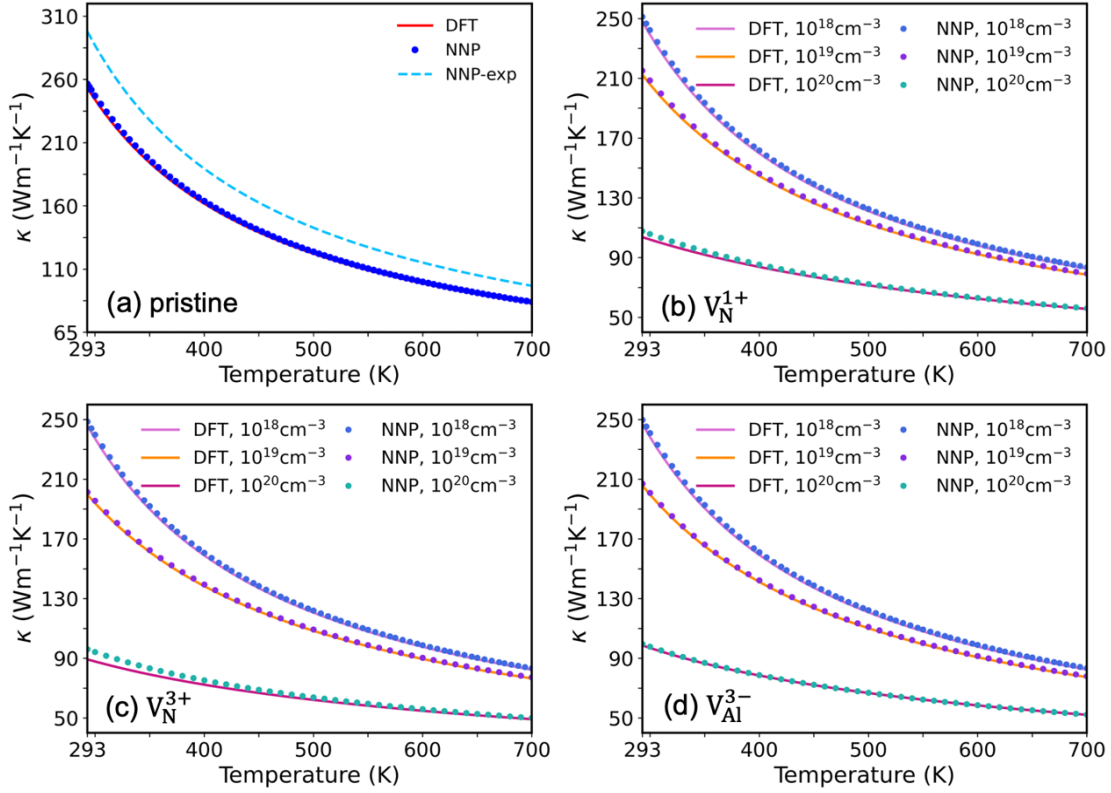

FIG. 6. Comparisons of thermal conductivities between predictions by NNP and DFT for (a) pristine AlN ($\tau_{anh}^{-1} + \tau_{iso}^{-1}$), and defect-laden AlN: (b) $\tau_{anh}^{-1} + \tau_{iso}^{-1} + \tau_{def}^{-1}(V_N^{1+})$, (c) $\tau_{anh}^{-1} + \tau_{iso}^{-1} + \tau_{def}^{-1}(V_N^{3+})$ and (d) $\tau_{anh}^{-1} + \tau_{iso}^{-1} + \tau_{def}^{-1}(V_{Al}^{3-})$. For pristine AlN, the thermal conductivities using the optimal and experimental lattice constants are shown. For defect-laden AlN, the average thermal conductivities using defect concentrations from $1 \times 10^{18}$ cm$^{-3}$ to $1 \times 10^{20}$ cm$^{-3}$ are presented.

In Figure 7, we further compare the effect of native defects ($V_N^{1+}$, $V_N^{3+}$ and $V_{Al}^{3-}$) on the thermal conductivity of AlN. When the defect concentration is as low as $10^{18}$ cm$^{-3}$, the magnitude of the effect of these three native impurities is minor and similar. When the defect concentration gets larger, the difference between those defects becomes more obvious. The effect of $V_N^{3+}$ on $\kappa$ is largest, $V_{Al}^{3-}$ is intermediate and $V_N^{1+}$ has the smallest effect, which is consistent with the comparisons of their phonon-defect scattering rates. This goes against the previous knowledge that $V_{Al}^{3-}$ might cause the largest effect on $\kappa$, and proves the significance of structural distortions induced by defects on phonon transport. It is worth noting that when using different calculation settings, such as a q-point mesh of 26×26×26 and cutoff radii of 5 Å and 5.4 Å (see Supplemental Material S7), this conclusion still holds.

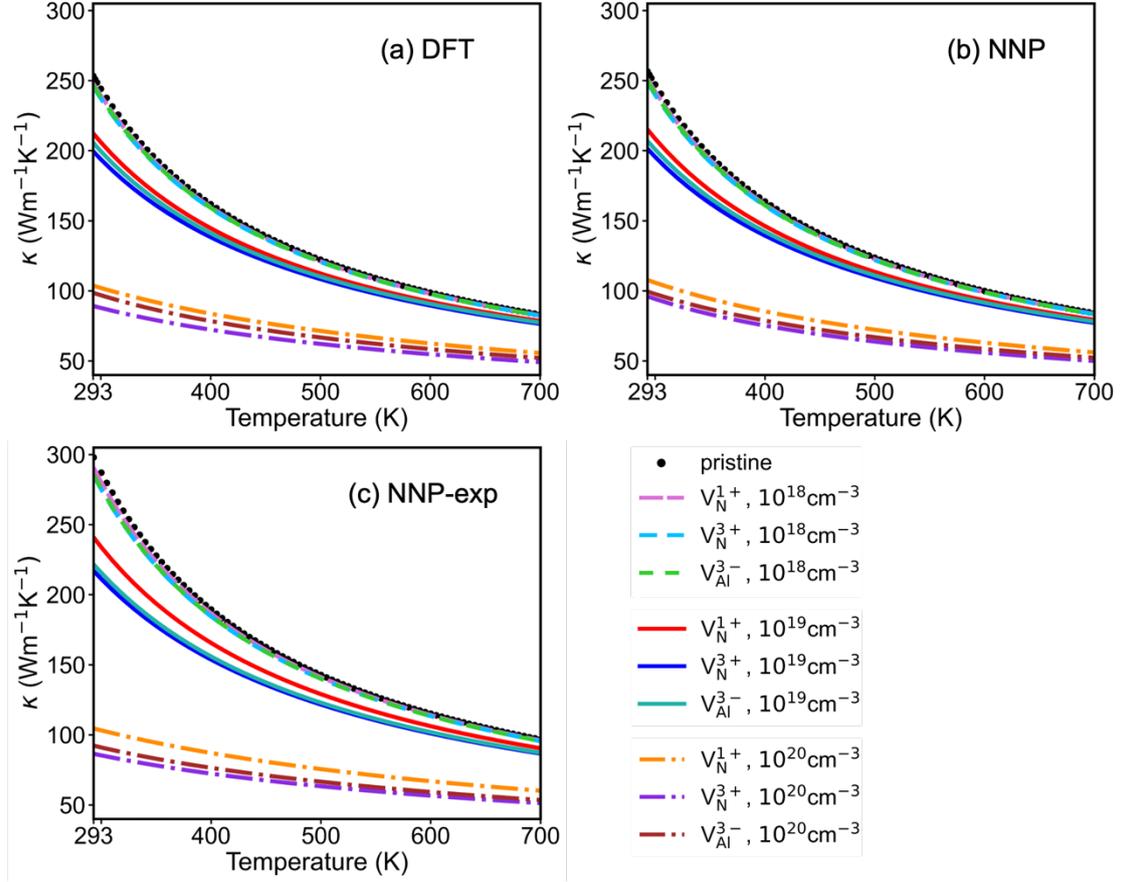

FIG. 7. Comparisons of thermal conductivities among pristine AlN ($\tau_{\text{anh}}^{-1} + \tau_{\text{iso}}^{-1}$), and defect-laden AlN: $\tau_{\text{anh}}^{-1} + \tau_{\text{iso}}^{-1} + \tau_{\text{def}}^{-1}(V_N^{1+})$, $\tau_{\text{anh}}^{-1} + \tau_{\text{iso}}^{-1} + \tau_{\text{def}}^{-1}(V_N^{3+})$ and $\tau_{\text{anh}}^{-1} + \tau_{\text{iso}}^{-1} + \tau_{\text{def}}^{-1}(V_{Al}^{3-})$ predicted using the optimal lattice constants by (a) DFT, (b) NNP and using the experimental lattice constants by (c) NNP. Defect concentrations of $1\times10^{18}$, $1\times10^{19}$ and $1\times10^{20}$ cm$^{-3}$ were considered.

## IV. CONCLUSION

In summary, we constructed a modified neural network potential (NNP) which could successfully describe the pristine and defect-laden AlN systems in various charge states. Its high accuracy was demonstrated in the predictions of various lattice dynamical properties, including phonon bands, phonon scattering rates of $\tau_{\text{anh}}^{-1}$, $\tau_{\text{iso}}^{-1}$ and $\tau_{\text{def}}^{-1}$, and thermal conductivity $\kappa$. We found that $V_N^{3+}$ reduces $\kappa$ most, followed by $V_{Al}^{3-}$, and $V_N^{1+}$ with the least impact. This was consistent with the comparisons between values of $\tau_{\text{def}}^{-1}$ and further confirmed by predictions via NNP together with the experimental lattice constants. Our study goes against, previous speculations that only considered the defect mass difference and reveals the importance for phonon scattering of structural distortions caused by the defects. Since no potentials had yet been developed to accurately depict the influence of defects with multiple charge states on phonon transport, our study additionally demonstrates that NNP can expand the high predictive capability of *ab initio* methods for phonon transport to large systems encompassing diverse charged defect species.

**ACKNOWLEDGMENTS**

This work was supported by JSPS KAKENHI Grants No. 19H02544, No. 19H04536, No. 21H05552, No. 23H04100, No. 24K01360. Some of the calculations used in this paper were performed using computer facilities at the ISSP Supercomputer Center and Information Technology Center, The University of Tokyo, and Institute for Materials Research, Tohoku University (MASAMUNE-IMR). This study was also supported by MCIN with funding from European Union NextGenerationEU (PRTR-C17.I1) promoted by the Government of Aragon.

# Reference

[1] S.-I. Tamura, Phys. Rev. B **27**, 858-866 (1983).
[2] S.-I. Tamura, Phys. Rev. B **30**, 849-854 (1984).
[3] W. Liu, A.A. Balandin, J. Appl. Phys. **97**, 073710 (2005).
[4] A. Sztein, J. Haberstroh, J.E. Bowers, S.P. DenBaars, S. Nakamura, J. Appl. Phys. **113**, 183707 (2013).
[5] R.L. Xu, M. Muñoz Rojo, S.M. Islam, A. Sood, B. Vareskic, A. Katre, N. Mingo, K.E. Goodson, H.G. Xing, D. Jena, E. Pop, J. Appl. Phys. **126**, 185105 (2019).
[6] C. Perez, A.J. McLeod, M.E. Chen, S.I. Yi, S. Vaziri, R. Hood, S.T. Ueda, X. Bao, M. Asheghi, W. Park, A.A. Talin, S. Kumar, E. Pop, A.C. Kummel, K.E. Goodson, ACS Nano **17**, 21240-21250 (2023).
[7] N.A. Katcho, J. Carrete, W. Li, N. Mingo, Physical Review B, **90** (2014) 094117.
[8] A. Katre, J. Carrete, T. Wang, G.K.H. Madsen, N. Mingo, Phys. Rev. Mater. **2**, 050602(R) (2018).
[9] Y. Dou, K. Shimizu, H. Fujioka, S. Watanabe, Comput. Mater. Sci. **244**, 113264 (2024).
[10] H. Babaei, R. Guo, A. Hashemi, S. Lee, Phys. Rev. Mater. **3**, 074603 (2019).
[11] K. Shimizu, Y. Dou, E.F. Arguelles, T. Moriya, E. Minamitani, S. Watanabe, Phys. Rev. B **106**, 054108 (2022).
[12] D. Dragoni, T.D. Daff, G. Csányi, N. Marzari, Phys. Rev. Mater. **2**, 013808 (2018).
[13] J. Behler, M. Parrinello, Phys. Rev. Lett. **98**, 146401 (2007).
[14] P. Giannozzi, S. Baroni, N. Bonini, M. Calandra, R. Car, C. Cavazzoni, D. Ceresoli, G.L. Chiarotti, M. Cococcioni, I. Dabo, A. Dal Corso, S. de Gironcoli, S. Fabris, G. Fratesi, R. Gebauer, U. Gerstmann, C. Gougoussis, A. Kokalj, M. Lazzeri, L. Martin-Samos, N. Marzari, F. Mauri, R. Mazzarello, S. Paolini, A. Pasquarello, L. Paulatto, C. Sbraccia, S. Scandolo, G. Sclauzero, A.P. Seitsonen, A. Smogunov, P. Umari, R.M. Wentzcovitch, J. Phys. Condens. Matter. **21**, 395502 (2009).
[15] J.P. Perdew, K. Burke, M. Ernzerhof, Phys. Rev. Lett. **77**, 3865-3868 (1996).
[16] K.F. Garrity, J.W. Bennett, K.M. Rabe, D. Vanderbilt, Comput. Mater. Sci. **81**, 446-452 (2014).
[17] S. Plimpton, J. Comput. Phys. **117**, 1-19 (1995).
[18] LAMMPS software, https://lammps.sandia.gov/.
[19] K. Shimizu, E.F. Arguelles, W. Li, Y. Ando, E. Minamitani, S. Watanabe, Phys. Rev. B **103**, 094112 (2021).
[20] B. Onat, E.D. Cubuk, B.D. Malone, E. Kaxiras, Phys. Rev. B **97**, 094106 (2018).
[21] E.D. Cubuk, B.D. Malone, B. Onat, A. Waterland, E. Kaxiras, J. Chem. Phys. **147**, 024104 (2017).
[22] J.L. Morales, J. Nocedal, ACM Trans. Math. Softw. **38**, 1-4 (2011).
[23] A. Togo, I. Tanaka, Scr. Mater. **108**, 1-5 (2015).
[24] W. Li, J. Carrete, N. A. Katcho, N. Mingo, Comput. Phys. Commun. **185**, 1747-1758 (2014).
[25] J. Carrete, B. Vermeersch, A. Katre, A. van Roekeghem, T. Wang, G.K.H. Madsen,

N. Mingo, Comput. Phys. Commun. **220**, 351-362 (2017).

[26] A. Katre, J. Carrete, N. Mingo, J. Mater. Chem. A **4**, 15940-15944 (2016).

[27] A. Katre, J. Carrete, B. Dongre, G.K.H. Madsen, N. Mingo, Phys. Rev. Lett. **119**, 075902 (2017).

[28] R.M. Pick, M.H. Cohen, R.M. Martin, Phys. Rev. B **1**, 910-920 (1970).

[29] D. Nilsson, E. Janzén, A. Kakanakova-Georgieva, J. Phys. D Appl. Phys. **49**, 175108 (2016).

[30] M. Schwoerer-Böhning, A.T. Macrander, J. Phys. Chem. Solids **61**, 485-487 (2000).

[31] F.J. Manjón, D. Errandonea, A.H. Romero, N. Garro, J. Serrano, M. Kuball, Phys. Rev. B **77**, 205204 (2008).

[32] Z. Cheng, Y.R. Koh, A. Mamun, J. Shi, T. Bai, K. Huynh, L. Yates, Z. Liu, R. Li, E. Lee, M.E. Liao, Y. Wang, H.M. Yu, M. Kushimoto, T. Luo, M.S. Goorsky, P.E. Hopkins, H. Amano, A. Khan, S. Graham, Phys. Rev. Mater. **4**, 044602 (2020).

[33] T.B. Boykin, G. Klimeck, Phys. Rev. B, **71**, 115215 (2005).

[34] T.B. Boykin, N. Kharche, G. Klimeck, M. Korkusinski, J. Phys. Condens. Matter **19**, 036203 (2007).

[35] P.B. Allen, T. Berlijn, D.A. Casavant, J.M. Soler, Phys. Rev. B **87**, 085322 (2013).

[36] https://gitlab.abinit.org/xuhe/unfolding.git.